Do school reforms shape study behavior at university? Evidence from an instructional time

reform

Do school reforms shape study efforts at university? Evidence from an instructional time reform*

Jakob Schwerter, Nicolai Netz, and Nicolas Hübner

Author Note

Jakob Schwerter, Center for Research on Education and School Development (IFS) & Faculty of Statistics, TU Dortmund University.

Nicolai Netz, German Centre for Higher Education Research and Science Studies (DZHW).

Nicolas Hübner, Institute of Education, University of Tübingen.

Correspondence concerning this manuscript should be directed to Jakob Schwerter, TU Dortmund University, Center for Research on Education and School Development (IFS), Martin-Schmeißer-Weg 13, 44227 Dortmund, Germany. Email: jakob.schwerter@tu-dortmund.de

* We thank Adam Ayaita, Morgan Polikoff, and Sebastian Lang for their valuable feedback on a prior version of this manuscript.

**Abstract**

Early-life environments can have long-lasting developmental effects. Interestingly, research on how school reforms affect later-life study behavior has hardly adopted this perspective. Therefore, we investigated a staggered school reform that reduced the number of school years and increased weekly instructional time for secondary school students in most German federal states. We analyzed this quasi-experiment in a difference-in-differences framework using representative large-scale survey data on 71,426 students who attended university between 1998 and 2016. We found negative effects of reform exposure on hours spent attending classes and on self-study, and a larger time gap between school completion and higher education entry. Our results support the view that research should examine unintended long-term effects of school reforms on individual life courses.

Keywords: School reform, difference-in-differences, higher education, study effort, long-term effect

**Do school reforms shape study efforts at university? Evidence from an instructional time reform**

There is compelling evidence that environments during childhood and adolescence influence a wide range of important outcomes in the later life course, such as skill development (Tymms et al., 2018), labor market performance (Gertler et al., 2014), and health (Taylor, 2010). Very few studies have investigated whether policy reforms that change the learning environment have long-lasting effects on students' attitudes towards study-related effort later in life, likely also because of challenges and cost of data collection. Such knowledge of potentially unintended long-term effects of policy interventions would be important for the planning, development, and implementation of future school reforms.

To address this research gap, we investigated whether a major school reform in Germany – known as the G8 reform – had long-term effects that remained perceptible during higher education. The G8 reform reduced the overall duration of lower secondary school from 9 to 8 years, while increasing weekly instructional time by about 3.68 hours or 12.5% (Hübner et al., 2022; Huebener et al., 2017), which led to reductions in students' leisure time (Hübner et al., 2017; Milde-Busch et al., 2010; Quis, 2018). The different German federal states introduced the G8 reform on a staggered basis, leading to exogenous variation in instructional time between states. First evidence showed that the reform increased students' school-related stress and decreased their health (Hübner et al., 2017; Marcus et al., 2020; Quis, 2018). Building on these findings, we tested whether the reform has also affected study behavior in terms of study efforts in higher education. Theoretically, students might have become accustomed to learning at a higher intensity in secondary school and maintained this intensity in higher education (habituation scenario). Conversely, they might also have reduced their learning intensity in higher education to recover from their stressful school years or because they have acquired a negative attitude toward formal education (compensation scenario). This question is of high importance, as several studies have shown a causal link between time investment in one's studies and performance in higher education (Andrietti & Velasco, 2015; Arulampalam et al., 2012; Bonesrønning & Opstad, 2015; Bratti & Staffolani, 2013; Ersoy, 2021; Grave, 2011; Metcalfe et al., 2019; Schwerter et al., 2022; Stinebrickner & Stinebrickner, 2008), which in turn improves labor market outcomes (Kittelsen Røberg & Helland, 2017). Therefore, our analysis of the long-term effects of secondary school reforms on study efforts at university closes an important research gap.

We exploited the quasi-experimental setting created by the staggered reform using a difference-in-differences (DiD) framework. To advance the literature in methodological terms, we applied the weighted group-time average treatment effects on the treated (ATT) proposed by Callaway and Sant'Anna (2020). This method overcomes several caveats of the two-way fixed effects (TWFE) method for examining time-staggered reforms. Our analysis was based on rich data on 71,426 university students in Germany.

**How Childhood Learning Environments Shape Individuals' Long-Term Behaviors: Insights from Different Academic Disciplines**

The sociological life course perspective highlights that experiences and decisions in the early life course create path dependencies that influence experiences and decisions later on (Elder et al., 2003). As a result of their socialization throughout childhood and adolescence, individuals tend to accumulate resources depending on their parents' economic, social, and cultural capital (Bourdieu, 1973). In this context, the life course approach illustrates that even slight differences in individuals' early-age resources and life chances may lead to severe social inequalities in later life (DiPrete & Eirich, 2006). Notably, such cumulative (dis)advantages may not only result from differences in individuals' early-age resource endowments; they are also shaped by the institutional settings that individuals pass through during the life course. Such institutional settings include the different stages of the education system, which may be abruptly altered by structural reforms (Mayer, 2004).

In line with these sociological propositions, psychological and school effectiveness research has produced a rich set of studies showing how early learning environments shape student learning and motivation later in life. For instance, prior studies have found that the negative association between a reference group's achievement and individuals' perception of their own abilities (the big-fish-little-pond effect) has effects that can still be detected five years after graduation from high school (Marsh & O'Mara, 2010). This line of inquiry has also produced evidence that being part of an effective learning environment at the beginning of school is substantially associated with educational attainment more than ten years later (Tymms et al., 2018). Further studies have suggested that stressful childhood environments are closely tied to health risks in adulthood (Taylor, 2010).

Relatedly, research on the economics of education has amply demonstrated how early childhood interventions or changes to schooling (environments) affect individuals' labor market outcomes later in life. For example, early childhood interventions that gave psychosocial stimulation to growth-restricted Jamaican toddlers substantially improved labor market outcomes 20 years later (Gertler et al., 2014). Abramitzky et al. (2021) further showed that economic incentives that changed children's schooling decisions in the early years of education increased later years of schooling and wages. Also, a free school choice program targeting disadvantaged students in Israel improved students' higher education enrollment in universities and teachers' colleges, as well as earnings later in life (Lavy, 2021). Moreover, changes in curriculum, such as intensified math instruction for low-ability students in lower secondary school, increased high school graduation rates and college enrolment (Cortes et al., 2015; Nomi & Raudenbush, 2016). Similarly, changing course selection opportunities in high school affected field of study choices in higher education (Biewen & Schwerter, 2022). Lastly, the literature has shown that more education (e.g., more years of compulsory schooling) lead to higher wages (Angrist & Krueger, 1991; Henderson et al., 2011; Oreopoulos, 2007), higher job prestige (Oreopoulos & Salvanes, 2011), and job satisfaction (Winkelmann & Winkelmann, 1998).

In summary, educational research from different disciplines illustrates that childhood experiences and learning environments early in the life course can have long-lasting influences. Therefore, changes in the structure of learning environments early on might also lead to changes in students' behavior much later in life. However, there are no studies that have tested this assumption. Hence, we investigated the effects of an instructional time reform (the G8 reform) on students' study efforts in higher education.

**Previous Research on Instructional Time Reforms**

A long-standing debate has addressed the importance of instructional time for student learning (Patall et al., 2010; Pischke, 2007). This debate has produced rich, albeit mixed evidence – with some studies suggesting negative effects of more instructional time, some suggesting positive effects, and others suggesting that the effects vary depending on student characteristics and examined outcomes (Allensworth et al., 2009; Andersen et al., 2016; Lavy, 2015; E. Meyer & Van Klaveren, 2013). Most of these studies have investigated whether increased instructional time led to higher student achievement. Theoretically, these studies are in line with traditional school effectiveness models assuming that properties of the learning process (e.g., time on task) influence how specific learning inputs are converted into specific learning outputs (Scheerens, 1990). They also align with theoretical frameworks such as the "Carroll Model of School Learning" (Carroll, 1988) or the concept of "Mastery Learning" (Bloom, 1968), which suggest that learning outcomes can be explained (and increased) by an optimal balance between students' aptitude and the amount of learning time.

In line with this are the results of a reform in Ontario, Canada, which shortened college-preparatory high school by one year. Krashinsky (2014) and Morin (2013) concluded that student achievement declined due to the reform, especially among lower-performing students. Thus, reducing instructional time alone led to a decline in achievement, in line with other findings that more years of schooling increase student achievement (Card, 1999; Lochner, 2011).

**Previous Research on the German G8 Reform**

In recent decades, most German federal states have introduced a reform compressing school time (Hübner et al., 2022). Unlike the reform in Canada, the German G8 reform not only shortened academic secondary school (*Gymnasium*) from 9 to 8 years, but also increased the amount of weekly instructional time (by about 3.68 hours or 12.5%  per week; Hübner et al., 2022) so that the total hours of instruction over the course of secondary school remained relatively constant. The G8 reform (i.e., "Gymnasium in eight years" reform) was one of the most fundamental reforms of the past decades and was accompanied by heated debates (Doersam & Lauber, 2019). Completing an academic-track school is still the most frequent way to fulfill the requirements for enrollment in a German university (Büchele, 2020). The *Gymnasium* is the highest of several secondary school tracks, which start after Grade 4 in most German federal states. Some states require a school track recommendation at the end Grade 4 from elementary school teachers, while others leave the decision to students' parents. Students who start secondary school in the *Realschule* (intermediate-track school) or *Hauptschule* (lower-track school) and perform well there have an opportunity to switch to an academic-track school after Grade 9 or 10, depending on the federal state (Biewen & Tapalaga, 2016).

The main concern of parents, teachers, and researchers about the reform was that increasing weekly instructional time could harm children's development (Kühn et al., 2013; Lehn, 2010; Marcus et al., 2020; Quis, 2018). Empirical evidence on the outcomes of the reform is mixed and varies across different outcomes:

*Cognitive abilities.* Andrietti and Su (2019), as well as Huebener et al. (2017), found that G8 students performed better on the PISA tests in Grade 9 than G9 students; however, students with low socio-economic status benefitted less or not at all compared to students with a higher socio-economic status. Further, G8 students had a higher probability of repeating a grade if they performed lower than average, which suggests that the increased weekly instructional time might have posed a particular challenge for low-performing students (Huebener et al., 2017).

Huebener and Marcus (2017) found that the increased grade repetition rates continued until the end of high school. Additionally, they found that G8 students had lower final grade point averages compared to G9 students, although their graduation rates were unchanged. Furthermore, as outlined by Hübner, Wagner, et al. (2017), G8 students exhibited lower performance in English reading and biology by the end of upper secondary school. Other studies have produced evidence that the lower GPA of G8 compared to G9 students was likely driven by worse grades in mathematics (Büttner & Thomsen, 2015). Interestingly, Dahmann (2017) found no significant differences between G8 and G9 students in (fluid and crystalline) intelligence by the end of high school.

*Stress and health.* Several studies have provided evidence that the G8 reform has increased school-related stress and health problems (Hübner et al., 2017, 2022; Marcus et al., 2020; Quis, 2018) and reduced emotional stability among upper-secondary students (Dahmann & Anger, 2014). Additionally, Hofmann and Mühlenweg (2018) showed that the G8 reform significantly reduced adolescents' self-rated mental health status.

*University performance, enrollment, and dropout.* Doersam and Lauber (2019), Marcus and Zambre (2019), Meyer et al. (2019), as well as Meyer and Thomsen (2016, 2018) more closely investigated the influence of the reform on long-term outcomes after school graduation. Doersam and Lauber (2019) found worse grades and a higher failure rate on exams among G8 students. Marcus and Zambre (2019) found that G8 students were more likely to delay enrollment, drop out, or change their fields of study in higher education. The finding on delayed higher education entry is consistent with Meyer et al. (2019) and Meyer and Thomsen (2016). However, Meyer & Thomsen (2018) and Doersam & Lauber (2019) did not find higher university dropout rates.

**Present Study**

Building on the aforementioned studies, we examined the effects of the German G8 reform, which increased instructional time in lower secondary school while reducing the overall years of schooling by one year in German academic track schools. We focused on reform-related changes in students' study efforts at university in terms of time spent attending classes and on self-study. Moreover, we investigated whether students invest their potentially freed-up time in paid work during their studies. Finally, less concentration on one's studies could also manifest in later enrollment in higher education due to the G8 reform. Therefore, we additionally analyzed if the reform increased the number of months between high school graduation and the start of one's first semester at university. This allows us to investigate whether students were more likely to take a gap year. Though our literature review showed that examining instructional time has a long history, we could not identify any studies that investigated its potential long-term impact on study efforts after school.

From a life course perspective (e.g., Becker & Schulze, 2013), it is theoretically plausible to assume that one of the following processes explains the effects of the examined policy reform: (a) habituation or (b) compensation. The habituation scenario (a) suggests that G8 students became used to a higher learning intensity in lower secondary school, which they then maintained in higher education. If this were the case, we would expect to find that G8 students spend more hours attending class and on self-study in higher education than G9 students. In contrast, the compensation scenario (b) suggests that the higher learning intensity in lower secondary school and associated reduced leisure time and health as well as increased stress created a desire among G8 students for more leisure time after school graduation. If this compensation scenario were true, we would expect to find that G8 students spent less hours attending class and on self-study and took more time between high school completion and university entry than G9 students.

The aforementioned literature on the impact of the G8 reform on achievement showed either no differences or even worse performance at the end of secondary school and university. Moreover, the cited health literature found negative effects of the reform. Therefore, scenario (a) seemed rather unlikely, and we expected to find evidence for scenario (b). Further support scenario (b), that G8 students had an increased desire for more leisure time during university, would be found if the hours spent on paid work during the semester were no higher among G8 students, as more time spent working during the semester would contradict the desire to experience more leisure time. G8 students had less free time during school, meaning that if scenario (b) were to hold true, students should not simply invest their freed-up time into more paid work. As additional support for scenario (b), G8 students should also take more time between graduation and university enrollment, as this would show that students were more likely to take a gap year after their more stressful time in secondary school.

We intentionally did not look at student performance, because grades assigned in similar courses at different universities are not based on standardized tests and are thus not comparable across universities.

**Method**

**Identification**

The start of the G8 reform differed between German federal states. The reform was implemented between 2007 and 2015 in all states but two (Table 1). Rheinland-Pfalz never implemented the reform, and in Schleswig-Holstein the reform was implemented one year after the last year covered by our dataset. Two other states, Sachsen and Thüringen, always had the G8 regime, as this was common in East Germany before reunification in 1990 and not changed afterwards. Therefore, we omitted these always-takers from the multivariate analysis. Similar to prior studies (Huebener et al., 2017), we also decided to drop Hessen because the reform was implemented gradually over the course of three years in this state and thus differed from the implementation in the other states. Moreover, due to the reduced number of years of schooling, the first G8 cohort graduated in the same year as the last G9 cohort. Hence, the first treatment cohort is the so-called "double cohort".

**Table 1:**
G8 Reform Introduction per State and Graduation Year

|  | 2007 | 2008 | 2009 | 2010 | 2011 | 2012 | 2013 | 2014 | 2015 |
|---|---|---|---|---|---|---|---|---|---|
| Baden-Württemberg | - | - | - | - | - | + | + | + | + |
| Bayern | - | - | - | - | + | + | + | + | + |
| Berlin | - | - | - | - | - | + | + | + | + |
| Brandenburg | - | - | - | - | - | + | + | + | + |
| Bremen | - | - | - | + | + | + | + | + | + |
| Hamburg | - | - | - | - | - | + | + | + | + |
| Hessen | - | - | - | - | - | - | + | + | + |
| Mecklenburg-Vorpommern | - | + | + | + | + | + | + | + | + |
| Niedersachsen | - | - | - | + | + | + | + | + | + |
| Nordrhein-Westfalen | - | - | - | - | - | - | + | + | + |
| Rheinland-Pfalz | - | - | - | - | - | - | - | - | - |
| Saarland | - | - | + | + | + | + | + | + | + |
| Sachsen | + | + | + | + | + | + | + | + | + |
| Sachsen-Anhalt | + | + | + | + | + | + | + | + | + |
| Schleswig-Holstein | - | - | - | - | - | - | - | - | - |
| Thüringen | + | + | + | + | + | + | + | + | + |

*Note:* - : Before implementation of G8 reform. + : After implementation of G8 reform. RP never implemented the reform. SH introduced the G8 reform in 2016, but we do not have data after 2015.

**Estimation Method**

The quasi-experimental difference-in-differences (DiD) method aims at estimating a treatment effect by comparing the effect of a quasi-experimental treatment on a treatment group with a baseline measurement and an untreated group. The resulting causal effect is commonly referred to as the ATT. The idea of DiD is to compare a treatment-induced change over time in the treatment group to a change without treatment in the control group, which can be written as:

$$\rho_{DiD} = (\bar{Y}_{s=1,t=2} - \bar{Y}_{s=1,t=1}) - (\bar{Y}_{d=0,t=2} - \bar{Y}_{d=0,t=1}),$$

where $s = 1$ refers to states in which students received the treatment at $t = 2$ (but not in $t = 1$), and $s = 0$ to students who did not receive the treatment at all. The underlying assumption is that the treatment group's trend would mimic the control group's trend if the former had not gotten the treatment. This means that the trend in both groups would be parallel in the absence of the reform (parallel trends assumption). Thus,

$$\bar{Y}^h_{s=1,t=2} = \bar{Y}_{s=1,t=1} + (\bar{Y}_{d=0,t=2} - \bar{Y}_{d=0,t=1}),$$

where $\bar{Y}^h_{s=1,t=2}$ is the hypothetical trend the treatment states would have had if they had received the reform. DiD is typically estimated in a regression framework using two-way fixed effects (TWFE) models (e.g., Biewen & Schwerter, 2022; Hackenberger et al., 2021; Henry et al., 2020; Jacob et al., 2017; Zimmer et al., 2017) to calculate valid standard errors. Such a regression model is formalized as follows:

$$y_{ist} = \alpha + \lambda_s + \lambda_t + \rho\, Treatment_{st} + X'_{ist}\beta + \varepsilon_{ist}$$

where index $i$ is on the individual level, $s$ is on an aggregated level like a state, and $t$ stands for the time point. Thus, $\lambda_s$ captures general state effects and $\lambda_t$ time effects. The variable *Treatment* is equal to 1 for the treatment group after the treatment took place and 0 otherwise. The vector $X$ includes all additional covariates, which may vary within individuals, states, and over time. The classical DiD setting is the 2×2 setting, in which there are two periods and two states. In the second period, one state gets the treatment, and the unobservable counterfactual development is interpolated using the untreated state.

In the case of the G8 reform, the different federal states had different years of implementation (Table 1). As outlined in more recent studies, using TWFE to estimate the ATT in such cases can be troublesome for several reasons (Callaway & Sant'Anna, 2020; applied in Gándara & Li, 2020). First, TWFE assume a constant treatment effect. However, TWFE suffer from heterogeneity bias if the treatment effect is not homogenous. In this case, one can show that the parallel trend bias is positive, while the heterogeneity bias is negative. Ex-ante, it is unclear which of these biases dominates or whether they cancel each other out. Consequently, the direction of the estimated coefficient in TWFE can be incorrectly estimated due to these biases (Callaway & Sant'Anna, 2020). Second, TWFE work well only in the classical 2×2 setting. If there are more time points with different treatment dates, TWFE aggregate all underlying 2×2 DiD treatment effects into one ATT. In doing so, TWFE put more weight on the underlying 2×2 DiDs at the center of the time distribution and those with more observations (see Callaway & Sant'Anna, 2020).

There are two methods to solve the TWFE treatment effect problem: weighted group-time ATT (Callaway & Sant'Anna, 2020; Sun & Abraham, 2021) and imputation (Athey & Imbens, 2021; Borusyak et al., 2022). Because we observe only two never-treated states with relatively few observations, we must base our comparison on states that have not yet been treated, which is easier with weighted group-time ATT (Callaway & Sant'Anna, 2020). Weighted group-time ATT are based on manually aggregated DiD models of each non-forbidden 2×2 DiD. Forbidden 2×2 DiD are those in which the already-treated are used as a comparison group for the to-be-treated. In other words, the main idea is to estimate the ATT of each 2×2 DiD and aggregate the aggregated treatment effect of interest. This model works well even if (i) treatment effects are heterogeneous in terms of time of adoption, (ii) treatment effects change over time, (iii) short-run effects are more pronounced than long-run effects, and (iv) treatment effect dynamics differ when people are first treated in a recession relative to economic expansion years. In this study, (i) is the case as the reform was implemented in a staggered way, because educational decisions are taken at the state level in Germany. Moreover, (ii) is likely as the share of students affected by the reform changed as each additional German state adopted the reform, meaning that the treatment effect may vary. Therefore, we rely on the R-package *did* (Callaway and Sant'Anna 2021) and use not-yet-treated individuals as the control group. Moreover, we use bootstrapped standard errors to correct for multiple testing. We provide statistical significance tests and confidence intervals at the 5% significance level. When including additional control variables, we use doubly robust estimations.

**DiD Assumptions**

For the DiD method to be valid, we must ensure that the reform did not affect the group composition. Concretely, we need to show that the reform did not induce students to choose a *Gymnasium* (i.e., academic school track) or other schools. Similarly, we need to check whether families moved to another state in order to evade the (staggered) reform introduction. Huebener and Marcus (2017) and Anger and Dahmann (2015) have critically examined and convincingly rejected these two potential pitfalls. We provide event-based graphs below to check the parallel trends assumption.

**Data**

To address our research question, we used data from the German Student Social Survey (https://doi.org/10.21249/DZHW:ssypool:1.0.1). Since 1951, this survey provides nationally representative data on the socio-demographic characteristics, previous educational history, and current situation of students in German higher education. Using a simple random sampling procedure, the survey addresses cross-sections of students at most public and private higher education institutions every three to four years in the summer semester (for details, see Apolinarski et al., 2021). We analyzed data from the survey years 2003, 2006, 2009, 2012, and 2016. We used the first two waves mainly for the pre-trend analyses, as no state had implemented the reform yet 2003 and 2006. Even though the survey is conducted only every three to four years, each high school graduating class is represented because university students in different years of study were surveyed. Thus, the 2009 survey cohort includes not only first-year students who graduated in 2008, but also more advanced students who graduated before 2008.

As the reform only applied to academic-track schools, the secondary school type with the highest academic standards in Germany, we only considered students who graduated from this track. Furthermore, we excluded non-traditional students to increase the homogeneity of our sample: We dropped students who had been enrolled in university for more than 16 semesters, for which the gap between high school graduation and survey year was larger than eight years, and students above age 30.

After applying these sample restrictions, our pooled dataset contained $N$ = 71,426 students, of which 59% were female (Table 2). Students in the sample had an average age of 23. The average number of semesters of university enrollment was 6.63. The vast majority (81%) attended a university and about one-fifth (19%) a university of applied sciences with a more career-oriented curriculum. Furthermore, 41% of students had attended school after the implementation of the reform (treatment group). Only a few students were born outside Germany (3%) and had children (2%).

**Table 2:**
Sample Description

|  | $N$ | Mean | SD | Min | Max |
|---|---|---|---|---|---|
| Treatment (1 = yes) | 71426 | 0.41 | 0.24 | 0 | 1 |
| HS graduation year | 71426 | 2008.27 | 5.14 | 1998 | 2016 |
| Female | 71426 | 0.59 | 0.49 | 0 | 1 |
| Age | 71426 | 22.99 | 2.35 | 16 | 30 |
| Semester of study | 71426 | 6.63 | 3.71 | 1 | 16 |
| University (vs. university of applied sciences) | 71426 | 0.81 | 0.39 | 0 | 1 |
| Child(ren) | 71426 | 0.02 | 0.12 | 0 | 1 |
| Immigrant background | 71426 | 0.03 | 0.18 | 0 | 1 |

*Note:* $N$ = number of observations, *SD* = standard deviation, *Min* = minimum, and *Max* = maximum (of possible values of each variable).

**Measures**

We investigated the effect of the G8 reform on eight outcome variables. These variables provide a comprehensive picture of students' study effort: First, we examined time spent attending classes, time spent on study-related activities, and working hours during a typical week during the term. The question used to capture these items was "During the last 'typical' week of the semester, how many hours a day did you spend on the following activities? Indicate this for each day of the week in hours. Please round to full hours!". We decided to aggregate the daily information into two distinct variables for workweek (MO-FR) and weekends (SA-SU). We also investigated a dichotomous variable that assessed whether students worked for pay during the semester. Lastly, we looked at the number of months between high school completion and university entry.

Table 3 shows that students spent about 12 to 13 hours on average attending class during the workweek and almost no time on the weekends. They further spent another 12 to 13 hours on self-study per week – about 9 hours during the workweek and 3 to 4 hours on the weekend. Moreover, they devoted about 3 to 4 hours during the workweek and 1 to 2 hours on the weekend to paid work. Around 64% of students worked during the semester. The average number of months between high school graduation and university entry was about 11 months.

**Table 3:**
Descriptive Statistics

|  | *N* | *Mean* | *SD* | *Min* | *Max* |
|---|---|---|---|---|---|
| *Workweek* |  |  |  |  |  |
| Hours spent attending class | 71426 | 12.41 | 11.49 | 0 | 71 |
| Hours spent on self-study | 71426 | 9.20 | 10.11 | 0 | 80 |
| Hours spent working | 71426 | 3.61 | 6.70 | 0 | 60 |
| *Weekends* |  |  |  |  |  |
| Hours spent attending class | 71426 | 0.16 | 1.06 | 0 | 24 |

|  | N | Mean | SD | Min | Max |
|---|---|---|---|---|---|
| Hours spent on self-study | 71426 | 3.48 | 4.48 | 0 | 38 |
| Hours spent working | 71426 | 1.22 | 3.18 | 0 | 32 |
| *Non-weekly variables* |  |  |  |  |  |
| Worked during the semester | 68756 | 0.64 | 0.48 | 0 | 1 |
| Months between high school completion and university entry | 71426 | 11.26 | 12.34 | 0 | 106 |

*Note:* $N$ = number of observations, $SD$ = standard deviation, $Min$ = minimum, and $Max$ = maximum (of possible values of each variable).

## Results

### Baseline Results

The baseline results in Table 4 indicated a significant G8 treatment effect on time spent attending classes during the workweek, hours spent on self-study during the weekend and on the weekend, and on the time gap between high school completion and university entry. According to these estimations, students affected by the G8 reform spent about 1.8 fewer hours attending class during the week. In addition, G8 students spent about 1.4 fewer hours during workweek and 0.5 fewer hours on the weekend on self-study. We also found a positive treatment effect on the duration between HS graduation and university entry of about 2 months. There was no statistically significant effect on time spent attending classes during the weekend, hours spent in paid work (during the workweek and the weekend), or the general probability of working. Overall, these results support the compensation scenario (b), which suggests that G8 students spent less time on self-study at university without increasing their working time.

**Table 4:**
Treatment Effects of the G8 Reform on Study Effort at University

| | ATT | SE | Lower CI | Upper CI |
|---|---|---|---|---|
| _Working days_ | | | | |
| Hours spent attending class | -1.7717* | 0.4835 | -3.0061 | -0.5373 |
| Hours spent on self-study | -1.4047* | 0.3668 | -2.3353 | -0.4740 |
| Hours spent working | -0.0619 | 0.1893 | -0.5679 | 0.4442 |
| _Weekends_ | | | | |
| Hours spent attending class | -0.0015 | 0.0267 | -0.0638 | 0.0607 |
| Hours spent on self-study | -0.5319* | 0.1927 | -1.0577 | -0.0061 |
| Hours spent working | -0.1685 | 0.1212 | -0.4793 | 0.1423 |
| _Non-weekly variables_ | | | | |
| Worked during the semester | 0.0413 | 0.0213 | -0.0151 | 0.0977 |
| Months between high school completion and university entry | 1.7379* | 0.3363 | 0.8671 | 2.6086 |

_Note:_ ATT refers to the weighted group-time average treatment effect on the treated following (Callaway & Sant'Anna, 2021). Standard errors (SE), lower and upper 95% confidence intervals (CI) are adjusted for multiple testing and calculated using bootstrapped standard errors. * $p < 0.05$

**Development of Treatment Effects Over Time Relative to the Year of Introduction**

We further decomposed the general ATT using an event-study design (Figure 1) in which all states' treatment effects are aggregated in terms of normalized time relative to the reform implementation. In so doing, we see (a) whether there is a pre-trend that violates the common trends assumption and (b) whether the ATT is constant for all post-treatment periods. This elucidates whether (a) the parallel trend assumption holds and (b) the treatment effect changes over time.

First, we did not observe any statistically significant differences in the pre-periods, supporting the common trends assumption for all examined outcome variables. For time spent attending classes, the effect was negative but not statistically significant when the treatment was first introduced. For the following two post-periods, we find negative and statistically significant treatment effects. Thereafter, the estimates were constant from post-periods 3 to 8, but the confidence intervals widen, resulting in statistically insignificant effects. Notably, the G8 reform was introduced between 2007 and 2013, and the last data collection point was in 2016. Thus, our data included only a few students from certain larger states with more than three post-periods. This explains the less efficient estimations for later periods.

The effect of the G8 reform on hours spent on self-study during workweek was significant for the double cohort and the next two graduation years. Again, the estimation results were relatively constant but less efficient for the later post-reform years. For hours spent on self-study on the weekend, we found a clear negative trend in the estimation up to the sixth post-period. Still, only the effect for the second post-period was statistically significant. Lastly, for the months between high school graduation and university entry, the effect was significant for the first to fifth post-periods. The graphs for time spent in class on the weekend and all three work-related outcomes did not reveal any additional information compared to Table 4, as there were no significant differences.

**Fig. 1:**
Event-study style results

Fig. 1a
Hours spent attending classes during workweek

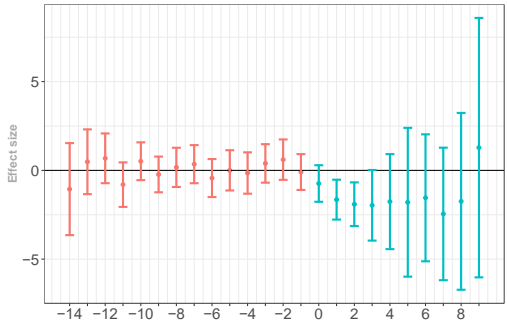

Fig. 1b
Hours spent attending classes during the weekend

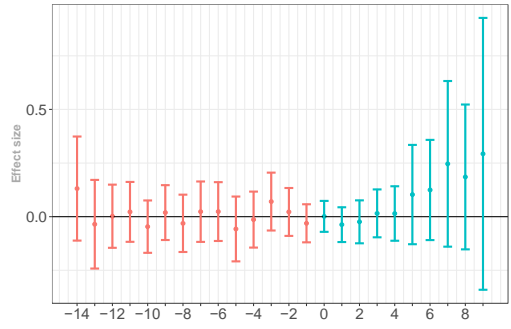

Fig. 1c
Hours spent on self-study during workweek

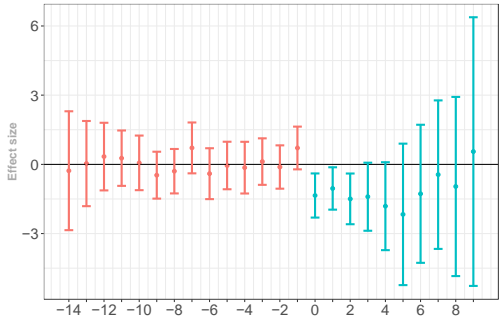

Fig. 1d
Hours spent on self-study during the weekend

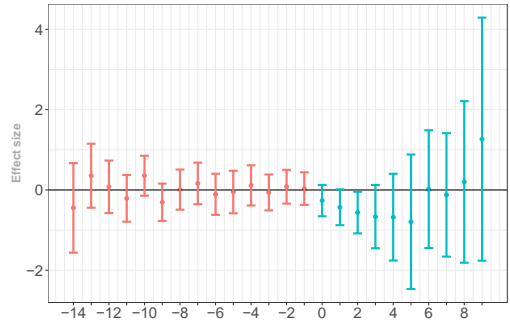

Fig. 1e
Hours spent working during workweek

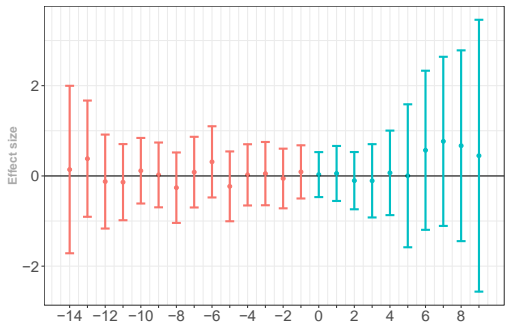

Fig. 1f
Hours spent working during the weekend

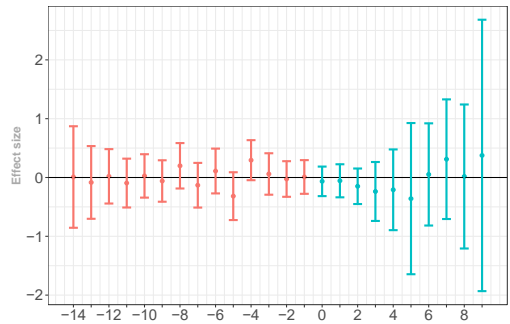

Fig. 1g
Working during the semester (yes/no)

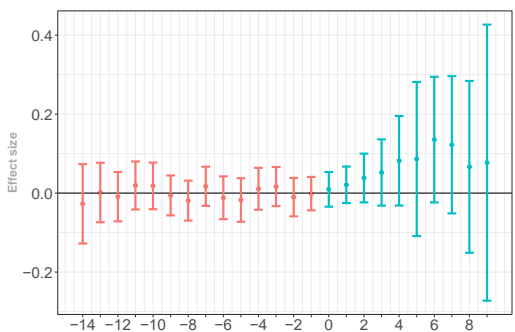

Fig. 1h
Months between high school completion and university entry

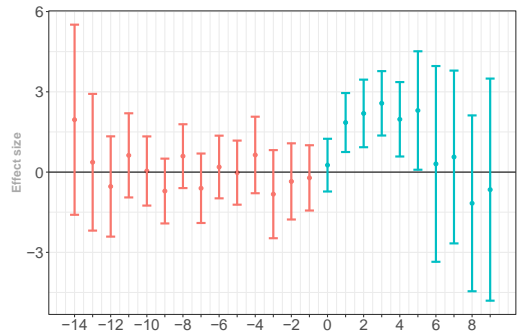

*Note*: The x-axis shows the time in years relative to the treatment. Negative values and red lines indicate measures before the treatment, and positive values and blue lines indicate measures after the treatment. The y-axis shows the effect of the treatment on the respective outcomes.

**Additional Control Variables**

Like a perfectly randomized experiment, estimates in a DiD framework should be robust to the inclusion of additional control variables. To check for differences between federal states, we included control variables for gender, age, and whether students attended a university or a university of applied sciences. For ease of interpretation, we compared the results presented in Table 4 graphically with the estimation results using additional control variables in Figure 3. We found no change in the interpretation of whether the treatment effect was statistically unequal to zero. Following Knol et al. (2011), we added 83.4% confidence intervals to examine whether the respective point estimates are statistically significantly different from each other at the 5% level. Except for the number of months between high school completion and university entry, all point estimates were similar. For the number of months between high school completion and university entry, the difference was only marginal. Hence, our previous estimation results were robust and not influenced by the included variables. This additionally supports the common trends assumption.

**Fig. 3:**
Comparing baseline results with estimations including additional control variables

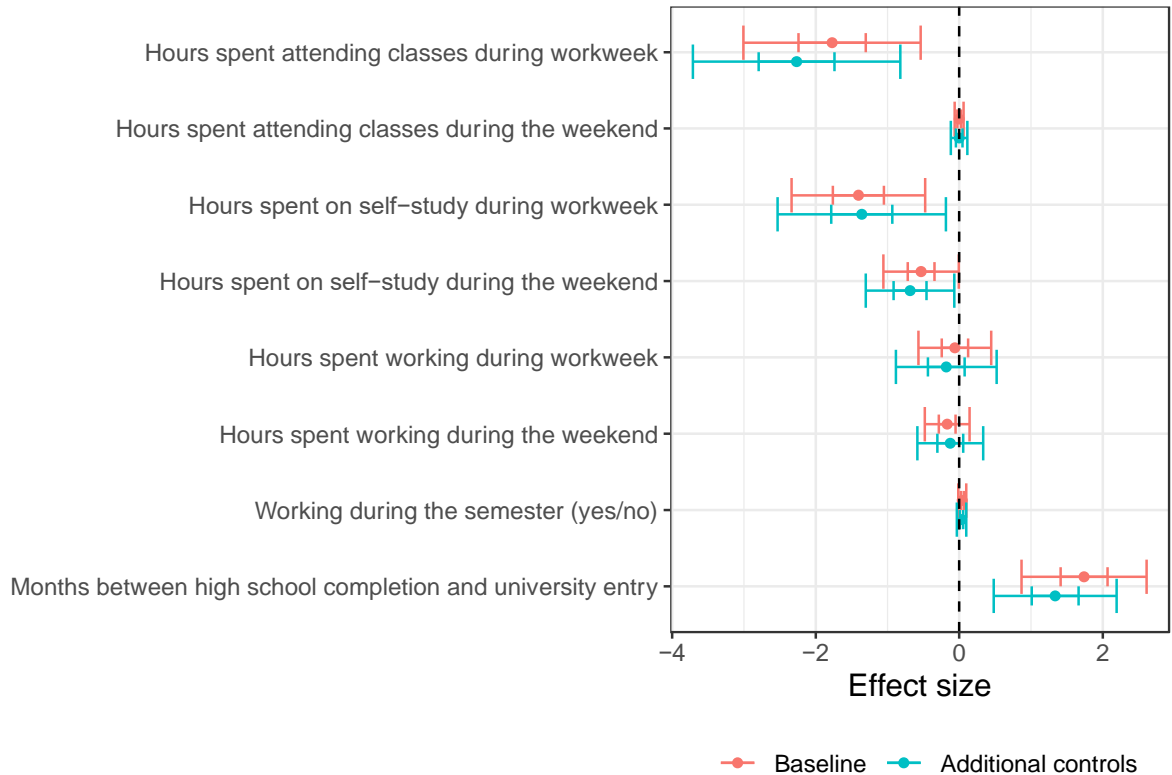

*Note*. We included the 95% and 83.4% confidence intervals for each estimate. We provide the 95% confidence interval to highlight whether an estimate is statistically different from zero at the 5% level. We also included the 83.4% confidence intervals to highlight whether the point estimates for the base regression (red) and the regression with additional control variables (blue) are significantly different from zero at the 5% level (Knol et al., 2011).

**Double cohort**

As the first treatment period also contained students from the old regime, we further present regression results in which the double cohort is treated as a pre-treatment cohort. Again, the results were relatively robust. Only for on self-study during the week did the upper confidence interval also include zero once we included additional control variables. Other than that, the results were not affected much by the coding of the double cohort. Therefore, these results are largely in line with the results presented in Fig1a to Fig1h.

**Fig 4:**
Double cohort as pre-treatment

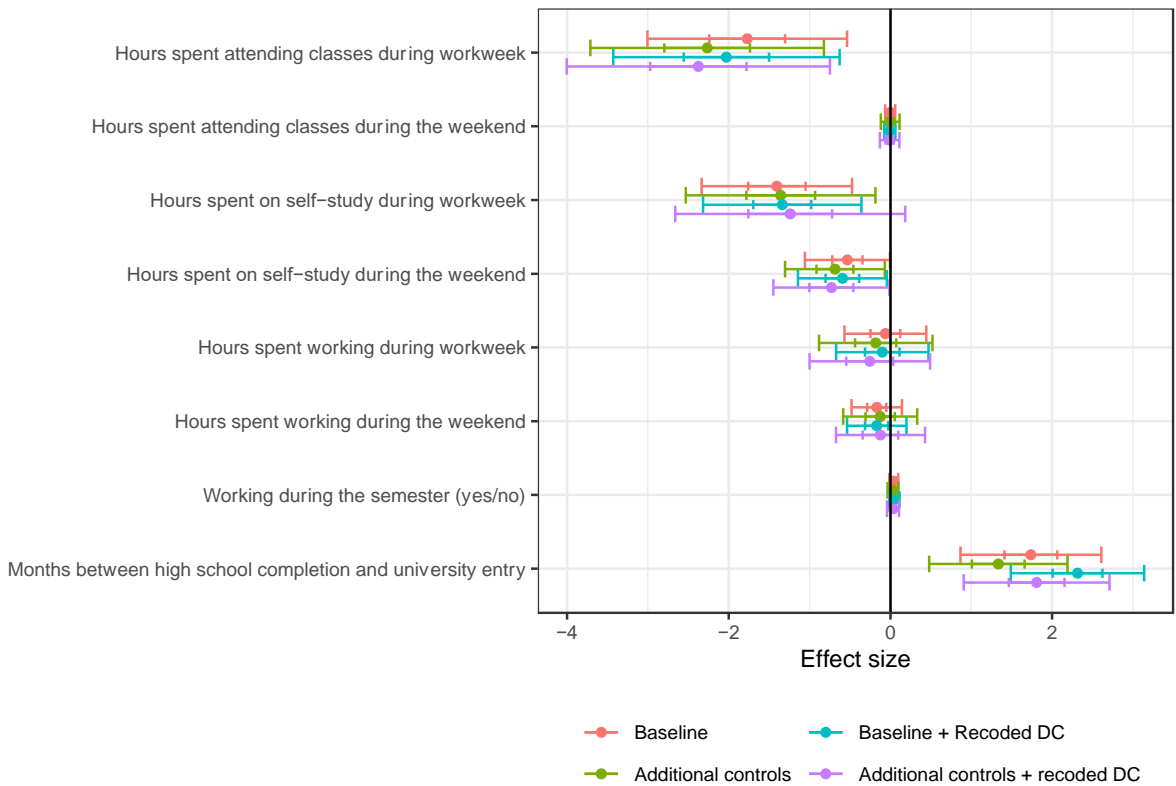

*Note*. We included the 95% and 83.4% confidence intervals for each estimate, like above in Fig. 3. In addition to the baseline regressions (red) and regressions including additional control variables (green), we added regressions in which the double cohort is recoded as pre-treatment cohort – once without additional control variables (blue), and once with (purple).

**Two-Way Fixed Effects Results**

This section demonstrates the necessity of using the *new* difference-in-differences method introduced by Callaway and Sant'Anna (2020). Typically, multiple specifications with different sets of control variables are included sequentially when using the TWFE. However, the biggest problem are *forbidden DiDs* in the TWFE, which compare always-takers or already-takers (students from states that already implemented the reform) to to-be-treated students. The following graphs showed the TWFE estimation for five different models. Following Huebener et al. (2017), the first TWFE model included the general DiD variables and a dummy for the double cohort (M1). The second model included the control variables for gender, age, semester of study, enrollment in a university vs. university of applied sciences, and having at least one child (M2). The third model added the type of degree students were pursuing (e.g. bachelor's or master's degree) (M3). The fourth model added a dummy variable for each field of study to M2 (M4). Lastly, M5 combines M3 and M4. Figure 4 presents the respective estimation results, including the results already presented in Figure 3, first using no additional control variables, and second including control variables, for easier comparison.

Figure 4a shows that the TWFE would have yielded estimates biased heavily towards zero – though still significant for time spent attending classes during the week. However, instead of an estimated reduction of almost two hours, the TWFE would estimate a treatment effect of about half an hour up to one hour.

For self-study on workweek (Figure 4c) and weekends (Figure 4d), the coefficients of the TWFE specifications are also biased towards zero and sometimes no longer statistically significant. This imprecise estimation highlights the problem with TWFE, as we would have missed the effect entirely. Only for the number of months between high school completion and university are the estimation results similar with both methods, albeit only once we included additional control variables.

**Fig 5:**

Comparing weighted group-time ATT with TWFE

Fig. 1a
Time spent attending classes during workweek

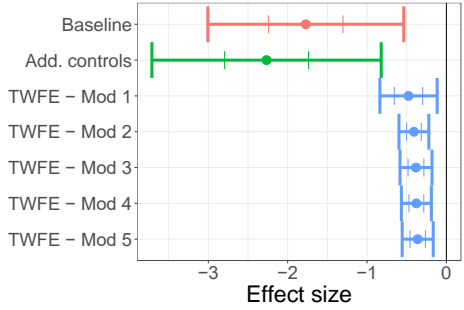

Fig. 1b
Time spent attending classes during the weekend

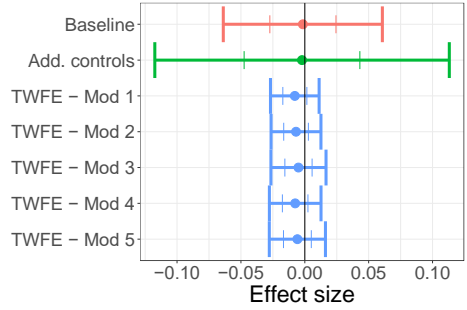

Fig. 1c
Hours spent on self-study during workweek

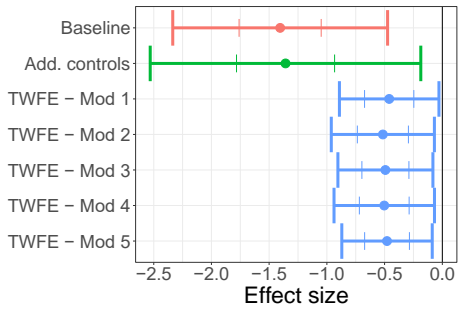

Fig. 1d
Hours spent on self-study during the weekend

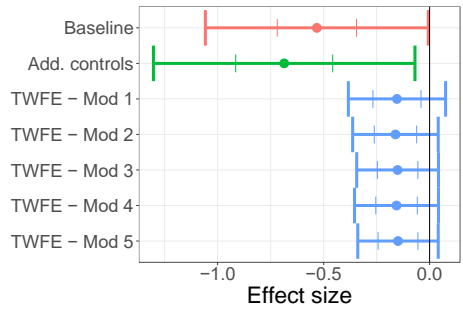

Fig. 1e
Hours spent working during workweek

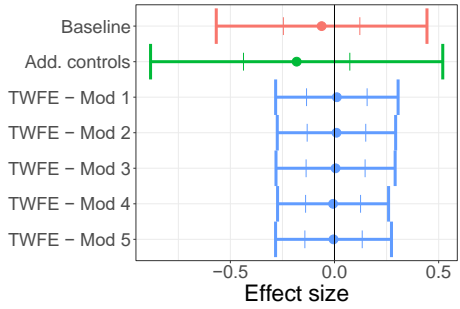

Fig. 1f
Hours spent working during the weekend

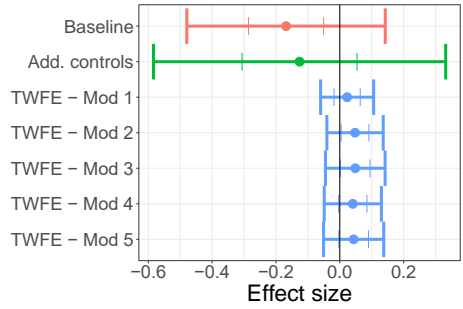

Fig. 1g
Working during the semester (yes/no)

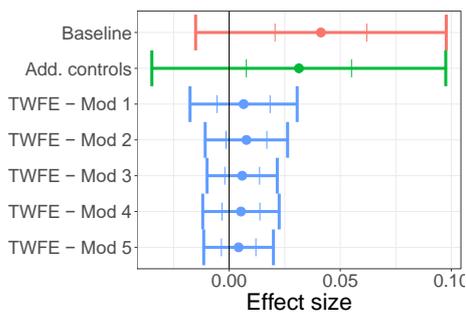

Fig. 1h
Months between high school completion and university entry

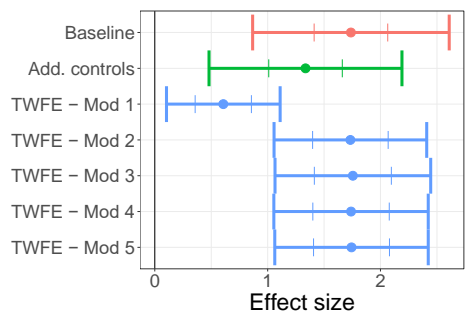

*Note.* The first two regression results show the weighted group average estimation (as above), while the other five regression (Mod 1 – Mod 5) show two-way fixed effects (TWFE) regressions with different sets of control variables.

## Discussion

Following a life course approach, we examined whether changes in the learning environment earlier in the life course have long-lasting effects on students' study effort later in life. Concretely, we investigated the effects of the G8 reform in Germany, which reduced secondary school by one year and increased average instructional time from 30-32 hours to 32-34 hours per week. Instructional time in the remaining eight years of secondary school was increased to compensate for the omitted year so that a similar amount of content could be covered. We tested whether the G8 reform influenced the duration between high school completion and university entry, as well as different measures of study effort at university.

Our results showed that G8 students spent less time attending class at university during the workweek than G9 students. The reform also reduced hours spent on self-study during the workweek and weekends. Thus, compressing students' time in school while increasing average weekly instructional time has detrimental effects on study effort at university. Furthermore, G8 students did not use their freed-up time to work more and were tended to enroll in university after a longer time gap compared to G9 students.

We started from two theoretically plausible scenarios explaining possible G8 reform effects: (a) habituation processes and (b) compensation processes. Our empirical results align with scenario (b): The estimated ATT suggest a compensation process, i.e., G8 students spend less time studying without working more. Thus, G8 students have more leisure time than G9 students. Moreover, the ATT were particularly pronounced in the first periods after the introduction of the treatment. After that, the treatment effect was reasonably stable, but not statistically significant due to large standard errors. One possible explanation for the lack of significance in the later post-reform periods is the relatively small or nonexistent number of G8 students from states that implemented the reform in 2012 or 2013. For example, from the states where the first G8 students graduated in 2013, there are only two cohorts of students who were affected by the reform in the data. Our estimations on working (hours) and the time gap between high school completion and university entry underline the relevance of scenario (b), as G8 students were more likely to take a gap year and less likely to work during the semester.

Our results are consistent with previous findings on the G8 reform: Several studies found that G8 students delayed their university entry, changed their fields of study more frequently, and were more likely to drop out of higher education (Marcus & Zambre, 2019; T. Meyer et al., 2019; T. Meyer & Thomsen, 2018). These patterns align with our finding that G8 students invest less time in their studies. G8 students seem to be less focused on their studies. Instead, they have more free time once they finish school: They enroll in higher education later than G9 students and invest less time in their studies once enrolled – without spending more time on paid work. These results underscore a new facet of a broadly visible pattern of the G8 reform having possibly detrimental effects.

While an increase in average performance was found among 9th graders using PISA data (Andrietti & Su, 2019; Huebener et al., 2017), the opposite was the case at the end of high school (Huebener & Marcus, 2017) and at university (Doersam & Lauber, 2019). However, there was no evidence of lower (fluid or crystalline) intelligence at the end of high school among G8 compared to G9 students (Dahmann, 2017). A possible explanation for why G8 students perform worse at university (Doersam & Lauber, 2019) despite not being less intelligent (Dörsam & Lauber, 2015) is relatively lower study effort we find in our analysis. This explanation is also very much in line with the literature showing that more effort leads to better educational performance in higher education (Andrietti & Velasco, 2015; Arulampalam et al., 2012; Bonesrønning & Opstad, 2015; Bratti & Staffolani, 2013; Ersoy, 2021; Grave, 2011; Metcalfe et al., 2019; Schwerter et al., 2022; Stinebrickner & Stinebrickner, 2008). Combining our results with those of the cited literature, we can thus conclude that students are overwhelmed by the increased instructional time in school, and hence (tend to) perform worse at university because they invest less time in their studies to compensate for having less free time during school. These results are particularly relevant given the crucial importance of lifelong learning. The G8 system does not seem to optimally prepare students for a world in which lifelong learning is a prerequisite for a successful life.

Rivkin & Schiman (2015) suggested that too many instructional hours per day increase the probability of fatigue and declining concentration. Thus, G8 students might just have been overwhelmed by the increased instructional time (Huebener et al., 2017). This fatigue and the feeling of being overwhelmed might be why these students are less focused at school. Other negative effects of the reform concern students' poorer health (Dahmann & Anger, 2014; Hofmann & Mühlenweg, 2018; Marcus et al., 2020; Quis & Reif, 2017). Students who experience fatigue, feel overwhelmed, experience more stress, and have overall poorer health while in school may reduce their learning load in higher education, when they have unrestricted control over it, to compensate for their stressful school years. The results regarding G8 students' health are in line with findings that one additional weekend day is not more important than shorter school days (Anderson & Walker, 2015).

In summary, empirical evidence seems to confirm the main concern of parents, teachers, and researchers that the more intensive daily instruction could have adverse effects on children's development in the long run (Kühn et al., 2013; Lehn, 2010). Future research could look into heterogenous effects of the G8 reform and the distributional composition of G8 and G9 students in a field of study, as well as the extent to which G8 and G9 students influenced each other in the first post-reform years, in which G8 and G9 students enrolled in university together.

## Limitations

While our results are robust to several sensitivity checks and align with previous results, our study also has limitations. First, the data on the number of hours spent studying are subjective, so minor bias is possible. However, we think it is unlikely that the reform changed students' response behavior, producing the differences we observe. Moreover, our data did not allow us to determine whether students only invested less time or whether they (also) obtained fewer ECTS credits per semester and thus made less progress in their studies. To date, there are no large-scale data suitable for such an analysis. Future research should thus also focus on university completion and examine whether G8 students are less likely to enroll in professional development training later in life.

Moreover, given that the DZHW surveys students only every three to four years, a higher university dropout rate for G8 students might theoretically bias the results towards zero, leading to lower-bound estimates. Since G8 students are more likely to drop out, the students who remained may have been more determined to follow through. However, students who dropped out probably invested less time attending class and on self-study, meaning that their early dropout likely led to an underestimation. Thus, our estimations are rather conservative.

## Conclusion

To our knowledge, no prior study has investigated the extent to which policy reforms implemented in lower secondary school create path dependencies that influence study effort in the later life course. We addressed this research gap by showing that a school reform increasing the weekly instructional time in lower secondary school for academic-track students in Germany decreased their time spent attending classes and on self-study at university. Our study also advances the corresponding methodological literature by presenting weighted group-time

ATT (Callaway & Sant'Anna, 2021), which overcome several problems of the TWFE in DiD frameworks. Our results suggest that researchers and policymakers should more seriously consider long-term effects when developing, implementing, and evaluating school reforms.

# References

Abramitzky, R., Lavy, V., & Pérez, S. (2021). The long-term spillover effects of changes in the return to schooling. *Journal of Public Economics*, *196*, 104369. https://doi.org/10.1016/j.jpubeco.2021.104369

Allensworth, E., Nomi, T., Montgomery, N., & Lee, V. E. (2009). College preparatory curriculum for all: Academic consequences of requiring algebra and english I for ninth graders in Chicago. *Educational Evaluation and Policy Analysis*, *31*(4), 367–391. https://doi.org/10.3102/0162373709343471

Andersen, S. C., Humlum, M. K., & Nandrup, A. B. (2016). Increasing instruction time in school does increase learning. *Proceedings of the National Academy of Sciences of the United States of America*, *113*(27), 7481–7484. https://doi.org/10.1073/pnas.1516686113

Anderson, D. M., & Walker, M. B. (2015). Does shortening the school week impact student performance? Evidence from the four-day school week. *Education Finance and Policy*, *10*(3), 314–349. https://doi.org/10.1162/EDFP_a_00165

Andrietti, V., & Su, X. (2019). The impact of schooling intensity on student learning: Evidence from a quasi-experiment. *Education Finance and Policy*, *15*(4), 679–701. https://doi.org/https://doi.org/10.1162/edfp_a_00263

Andrietti, V., & Velasco, C. (2015). Lecture Attendance, Study Time, and Academic Performance: A Panel Data Study. *Journal of Economic Education*, *46*(3), 239–259. https://doi.org/10.1080/00220485.2015.1040182

Angrist, J. D., & Krueger, A. B. (1991). Does Compulsory School Attendance Affect Schooling and Earnings? *Quarterly Journal of Economics*, *106*(4), 979–1014. https://doi.org/10.1017/CBO9781107415324.004

Apolinarski, B., Becker, K., Borchert, L., Bornkessel, P., Brandt, T., Fabian, G., Heißenberg, S., Isserstedt, W., Kandulla, M., Leszczensky, M., Link, J., Middendorff, E., Netz, N., Poskowsky, J., Schnitzer, K., Weber, S., & Wolter, A. (2021). *17th – 21st Social Survey 2003 – 2016. Data Collection 2003, 2009, 2012, 2016. Version: 1.0.0. Data Package Access Way: Download-SUF* (E. Middendorff & M. Wallis (eds.)). https://doi.org/10.21249/DZHW:ssypool:1.0.1

Arulampalam, W., Naylor, R. A., & Smith, J. (2012). Am I missing something? The effects of absence from class on student performance. *Economics of Education Review*, *31*(4), 363–375. https://doi.org/10.1016/j.econedurev.2011.12.002

Athey, S., & Imbens, G. W. (2021). Design-based analysis in Difference-In-Differences settings with staggered adoption. *Journal of Econometrics*, *226*(1), 62–79. https://doi.org/10.1016/j.jeconom.2020.10.012

Becker, R., & Schulze, A. (2013). *Bildungskontexte. Strukturelle Voraussetzungen und Ursachen ungleicher Bildungschancen* (R. Becker & A. Schulze (eds.)). Springer. https://doi.org/10.1007/978-3-531-18985-7

Biewen, M., & Schwerter, J. (2022). Does more maths and natural sciences in high school increase the share of female STEM workers? Evidence from a curriculum reform. *Applied Economics*, *54*(16), 1889–1911. https://doi.org/10.1080/00036846.2021.1983139

Biewen, M., & Tapalaga, M. (2016). Life-cycle educational choices in a system with early tracking and "second chance" options. *Economics of Education Review*. https://doi.org/10.1016/j.econedurev.2016.11.008

Bloom, B. S. (1968). Learning for Mastery. *Evaluation Comment*, *1*(2), 1854–1854. https://doi.org/10.1007/978-1-4419-1428-6_4645

Bonesrønning, H., & Opstad, L. (2015). Can student effort be manipulated? Does it matter? *Applied Economics*, *47*(15), 1511–1524. https://doi.org/10.1080/00036846.2014.997923

Borusyak, K., Jaravel, X., & Spiess, J. (2022). *Revisiting Event Study Designs: Robust and Efficient Estimation*. 1–54. http://arxiv.org/abs/2108.12419

Bourdieu, P. (1973). Cultural Reproduction and Social Reproduction, Knowledge, Education and Social Change. In R. Brown (Ed.), *Papers in the Sociology of Education* (pp. 71–112). Tavistock Publication.

Bratti, M., & Staffolani, S. (2013). Student Time Allocation and Educational Production Functions. *Annals of Economics and Statistics*, *111/112*, 103. https://doi.org/10.2307/23646328

Büchele, S. (2020). Should we trust math preparatory courses? An empirical analysis on the impact of students' participation and attendance on short- and medium-term effects. *Economic Analysis and Policy*, *66*, 154–167. https://doi.org/10.1016/j.eap.2020.04.002

Büttner, B., & Thomsen, S. L. (2015). Are We Spending Too Many Years in School? Causal Evidence of the Impact of Shortening Secondary School Duration. *German Economic Review*, *16*(1), 65–86. https://doi.org/10.1111/geer.12038

Callaway, B., & Sant'Anna, P. H. C. (2021). Difference-in-Differences with multiple time periods. *Journal of Econometrics*, *225*(2), 200–300. https://doi.org/10.1016/j.jeconom.2020.12.001

Card, D. (1999). The causal effect of education on earnings. In D. C. Orley C. Ashenfelter (Ed.), *Handbook of Labor Economics* (3rd ed., pp. 1801–1863). Elsevier Science. https://doi.org/10.1016/S1573-4463(99)03011-4

Carroll, J. B. (1988). A 25-Year Retrospective and Prospective View. *Educational Research*, *18*(1), 26–31.

Cortes, K. E., Goodman, J. S., & Nomi, T. (2015). Intensive math instruction and educational attainment: Long-run impacts of double-dose algebra. *Journal of Human Resources*, *50*(1), 108–158. https://doi.org/10.3368/jhr.50.1.108

Dahmann, S. C. (2017). How does education improve cognitive skills? Instructional time versus timing of instruction. *Labour Economics*, *47*, 35–47. https://doi.org/10.1016/j.labeco.2017.04.008

Dahmann, S. C., & Anger, S. (2014). The Impact of Education on Personality: Evidence from a German High School Reform. *SSRN Electronic Journal*, *8139*. https://doi.org/10.2139/ssrn.2432423

DiPrete, T. A., & Eirich, G. M. (2006). Cumulative advantage as a mechanism for inequality: A review of theoretical and empirical developments. *Annual Review of Sociology*, *32*, 271–297. https://doi.org/10.1146/annurev.soc.32.061604.123127

Doersam, M., & Lauber, V. (2019). The Effect of a Compressed High School Curriculum on University Performance. *Working Paper Series of the Department of Economics, University Konstanz*, *2019–03*. https://ideas.repec.org/p/knz/dpteco/1903.html

Dörsam, M., & Lauber, V. (2015). The Effect of a Compressed High School Curriculum on University Performance Beiträge. *Conference Paper*.

Elder, G., Johnson, M., & Crosnoe, R. (2003). The Emergence and Development of Life Course Theory. In *Handbook of the Life Course* (pp. 3–19). Springer. https://doi.org/https://doi.org/10.1007/978-0-306-48247-2_1

Ersoy, F. (2021). Returns to effort: experimental evidence from an online language platform. *Experimental Economics*, *24*(3), 1047–1073. https://doi.org/10.1007/s10683-020-09689-1

Gándara, D., & Li, A. (2020). Promise for Whom? "Free-College" Programs and Enrollments by Race and Gender Classifications at Public, 2-Year Colleges. *Educational Evaluation and Policy Analysis*, *42*(4), 603–627. https://doi.org/10.3102/0162373720962472

Gertler, P., Heckman, J., Pinto, R., Zanolini, A., Vermeersch, C., Walker, S., Chang, S. M., & Grantham-McGregor, S. (2014). Labor market returns to an early childhood stimulation intervention in Jamaica. *Science*, *344*(6187), 998–1002.

Grave, B. S. (2011). The effect of student time allocation on academic achievement. *Education Economics*, *19*(3), 291–310. https://doi.org/10.1080/09645292.2011.585794

Hackenberger, A., Rümmele, M., Schwerter, J., & Sturm, M. (2021). Elections and unemployment benefits for families: Did the Family Benefit Dispute affect election outcomes in Germany? *European Journal of Political Economy*, *66*, 101955. https://doi.org/10.1016/j.ejpoleco.2020.101955

Henderson, D. J., Polachek, S. W., & Wang, L. (2011). Heterogeneity in schooling rates of return. *Economics of Education Review*, *30*(6), 1202–1214. https://doi.org/10.1016/j.econedurev.2011.05.002

Henry, G. T., Pham, L. D., Kho, A., & Zimmer, R. (2020). Peeking Into the Black Box of School Turnaround: A Formal Test of Mediators and Suppressors. *Educational Evaluation and Policy Analysis*, *42*(2), 232–256. https://doi.org/10.3102/0162373720908600

Hofmann, S., & Mühlenweg, A. (2018). Learning intensity effects in students' mental and physical health – Evidence from a large scale natural experiment in Germany. *Economics of Education Review*, *67*, 216–234. https://doi.org/10.1016/j.econedurev.2018.10.001

Hübner, N., Wagner, W., Kramer, J., Nagengast, B., & Trautwein, U. (2017). Die G8-Reform in Baden-Württemberg: Kompetenzen, Wohlbefinden und Freizeitverhalten vor und nach der Reform. *Zeitschrift Für Erziehungswissenschaft*, *20*(4), 748–771. https://doi.org/10.1007/s11618-017-0737-3

Hübner, N., Wagner, W., Meyer, J., & Watt, H. M. G. (2022). To Those Who Have, More Will Be Given ? Effects of an Instructional Time Reform on Gender Disparities in STEM Subjects, Stress, and Health. *Frontiers in Psychology*, *13*, 1–15. https://doi.org/10.3389/fpsyg.2022.816358

Huebener, M., Kuger, S., & Marcus, J. (2017). Increased instruction hours and the widening gap in student performance. *Labour Economics*, *47*, 15–34. https://doi.org/10.1016/j.labeco.2017.04.007

Huebener, M., & Marcus, J. (2017). Compressing instruction time into fewer years of schooling and the impact on student performance. *Economics of Education Review*, *58*, 1–14. https://doi.org/10.1016/j.econedurev.2017.03.003

Jacob, B., Dynarski, S., Frank, K., & Schneider, B. (2017). Are Expectations Alone Enough? Estimating the Effect of a Mandatory College-Prep Curriculum in Michigan. *Educational Evaluation and Policy Analysis*, *39*(2), 333–360. https://doi.org/10.3102/0162373716685823

Kittelsen Røberg, K. I., & Helland, H. (2017). Do grades in higher education matter for labour market rewards? A multilevel analysis of all Norwegian graduates in the period 1990–2006. *Journal of Education and Work*, *30*(4), 383–402. https://doi.org/10.1080/13639080.2016.1187265

Knol, M. J., Pestman, W. R., & Grobbee, D. E. (2011). The (mis)use of overlap of confidence intervals to assess effect modification. *European Journal of Epidemiology*, *26*(4), 253–254. https://doi.org/10.1007/s10654-011-9563-8

Krashinsky, H. (2014). How would one extra year of high school affect academic performance in university? Evidence from an educational policy change. *Canadian Journal of Economics*, *47*(1), 70–97. https://doi.org/10.1111/caje.12066

Kühn, S. M., Van Ackeren, I., Bellenberg, G., Reintjes, C., & Im Brahm, G. (2013). Wie viele Schuljahre bis zum Abitur?: Eine multiperspektivische Standortbestimmung im Kontext der aktuellen Schulzeitdebatte. *Zeitschrift Fur Erziehungswissenschaft*, *16*(1), 115–136. https://doi.org/10.1007/s11618-013-0339-7

Lavy, V. (2015). Do Differences in Schools' Instruction Time Explain International Achievement Gaps? Evidence from Developed and Developing Countries. *Economic Journal*, *125*(588), 397–424. https://doi.org/10.1111/ecoj.12233

Lavy, V. (2021). The Long-Term Consequences of Free School Choice. *Journal of the European Economic Association*, *19*(3), 1734–1781. https://doi.org/10.1093/jeea/jvab001

Lehn, B. vom. (2010). *Generation G8: wie die Turbo-Schule Schüler und Familien ruiniert*. Beltz.

Lochner, L. (2011). Nonproduction Benefits of Education: Crime, Health, and Good Citizenship. In L. W. Eric A. Hanushek, Stephen J. Machin (Ed.), *Handbook of the Economics of Education* (Vol. 4, pp. 183–282). Elsevier. https://doi.org/10.1016/B978-0-444-53444-6.00002-X

Marcus, J., Reif, S., Wuppermann, A., & Rouche, A. (2020). Increased instruction time and stress-related health problems among school children. *Journal of Health Economics*, *70*. https://doi.org/10.1016/j.jhealeco.2019.102256

Marcus, J., & Zambre, V. (2019). The effect of increasing education efficiency on university enrollment: Evidence from administrative data and an unusual schooling reform in Germany. *Journal of Human Resources*, *54*(2), 468–502. https://doi.org/10.3368/jhr.54.2.1016.8324R

Marsh, H. W., & O'Mara, A. J. (2010). Long-term total negative effects of school-average ability on diverse educational outcomes: Direct and indirect effects of the big-fish-little-pond effect. *Zeitschrift Fur Padagogische Psychologie*, *24*(1), 51–72. https://doi.org/10.1024/1010-0652/a000004

Mayer, K. U. (2004). Whose Lives? How History, Societies, and Institutions Define and Shape Life Courses. *Research in Human Development*, *1*(3), 161–187. https://doi.org/10.1207/s15427617rhd0103_3

Metcalfe, R., Burgess, S., & Proud, S. (2019). Students' effort and educational achievement: Using the timing of the World Cup to vary the value of leisure. *Journal of Public Economics*, *172*, 111–126. https://doi.org/10.1016/j.jpubeco.2018.12.006

Meyer, E., & Van Klaveren, C. (2013). The effectiveness of extended day programs: Evidence from a randomized field experiment in the Netherlands. *Economics of Education Review*, *36*, 1–11.

https://doi.org/10.1016/j.econedurev.2013.04.002

Meyer, T., & Thomsen, S. L. (2016). How important is secondary school duration for postsecondary education decisions? Evidence from a natural experiment. *Journal of Human Capital*, *10*(1), 67–108. https://doi.org/10.1086/684017

Meyer, T., & Thomsen, S. L. (2018). The role of high-school duration for university students' motivation, abilities and achievements. *Education Economics*, *26*(1), 24–45. https://doi.org/10.1080/09645292.2017.1351525

Meyer, T., Thomsen, S. L., & Schneider, H. (2019). New Evidence on the Effects of the Shortened School Duration in the German States: An Evaluation of Post-secondary Education Decisions. *German Economic Review*, *20*(4), 201–253. https://doi.org/10.1111/geer.12162

Milde-Busch, A., Blaschek, A., Borggräfe, I., von Kries, R., Straube, A., & Heinen, F. (2010). Is There an Association between the Reduced School Years in Grammar Schools and Headache and Other Health Complaints in Adolescent Students? *Klinische Pädiatrie*, *222*(4), 255–260. https://dx.doi.org/10.1055/s-0030-1252012

Morin, L. P. (2013). Estimating the benefit of high school for university-bound students: Evidence of subject-specific human capital accumulation. *Canadian Journal of Economics*, *46*(2), 441–468. https://doi.org/10.1111/caje.12019

Nomi, T., & Raudenbush, S. W. (2016). Making a Success of "Algebra for All": The Impact of Extended Instructional Time and Classroom Peer Skill in Chicago. *Educational Evaluation and Policy Analysis*, *38*(2), 431–451. https://doi.org/10.3102/0162373716643756

Oreopoulos, P. (2007). Do dropouts drop out too soon? Wealth, health and happiness from compulsory schooling. *Journal of Public Economics*, *91*(11–12), 2213–2229. https://doi.org/https://doi.org/10.1016/j.jpubeco.2007.02.002

Oreopoulos, P., & Salvanes, K. G. (2011). Priceless: The Nonpecuniary Benefits of Schooling. *Journal of Economic Perspective*, *25*(1), 159–184. https://doi.org/10.1257/jep.25.1.159

Patall, E. A., Cooper, H., & Allen, A. B. (2010). Extending the school day or school year: A systematic review of research (1985-2009). *Review of Educational Research*, *80*(3), 401–436. https://doi.org/10.3102/0034654310377086

Pischke, J.-S. (2007). The Impact Of Length Of The School Year On Student Performance And Earnings: Evidence From The German Short School Years. *The Economic Journal*, *117*(523), 1216–1242. https://doi.org/https://doi.org/10.1111/j.1468-0297.2007.02080.x

Quis, J. S. (2018). Does Compressing High School Duration Affect Students' Stress and Mental Health? Evidence from the National Educational Panel Study. *Journal of Economics and Statistics*, *238*(5), 441–476. https://doi.org/https://doi.org/10.1515/jbnst-2018-0004

Quis, J. S., & Reif, S. (2017). Health Effects of Instruction Intensity: Evidence from a Natural Experiment in German High-Schools. *SSRN Electronic Journal*. https://doi.org/10.2139/ssrn.3051700

Rivkin, S. G., & Schiman, J. C. (2015). Instruction Time, Classroom Quality, and Academic Achievement. *Economic Journal*, *125*(November), 425–448. https://doi.org/10.1111/ecoj.12315

Scheerens, J. (1990). School Effectiveness Research and the Development of Process Indicators of School Functioning. *School Effectiveness and School Improvement*, *1*(1), 61–80. https://doi.org/10.1080/0924345900010106

Schwerter, J., Dimpfl, T., Bleher, J., & Murayama, K. (2022). Benefits of additional online practice opportunities in higher education. *Internet and Higher Education*, *53*, 100834. https://doi.org/https://doi.org/10.1016/j.iheduc.2021.100834

Stinebrickner, R., & Stinebrickner, T. R. (2008). The causal effect of studying on academic performance. *B.E. Journal of Economic Analysis and Policy*, *8*(1). https://doi.org/10.2202/1935-1682.1868

Sun, L., & Abraham, S. (2021). Estimating dynamic treatment effects in event studies with heterogeneous treatment effects. *Journal of Econometrics*, *225*(2), 175–199. https://doi.org/https://doi.org/10.1016/j.jeconom.2020.09.006

Taylor, S. E. (2010). Mechanisms linking early life stress to adult health outcomes. *Proceedings of the National Academy of Sciences of the United States of America*, *107*(19), 8507–8512. https://doi.org/10.1073/pnas.1003890107

Tymms, P., Merrell, C., & Bailey, K. (2018). The long-term impact of effective teaching. *School Effectiveness and School Improvement*, *29*(2), 242–261. https://doi.org/10.1080/09243453.2017.1404478

Winkelmann, L., & Winkelmann, R. (1998). Why Are the Unemployed So Unhappy? Evidence from Panel Data. *Economica*, *65*, 1–15. https://doi.org/10.1111/1468-0335.00111

Zimmer, R., Henry, G. T., & Kho, A. (2017). The Effects of School Turnaround in Tennessee's Achievement School District and Innovation Zones. *Educational Evaluation and Policy Analysis*, *39*(4), 670–696. https://doi.org/10.3102/0162373717705729